\documentclass
[aps,prd,twocolumn,showpacs,nofootinbib,amsmath,amssymb,floatfix,superscriptaddress,showkeys]{revtex4-1}
\usepackage[colorlinks, linkcolor=blue, anchorcolor=blue, citecolor=blue, colorlinks=true, urlcolor=blue]{hyperref}
\usepackage{amssymb}
\usepackage{amsmath}
\usepackage{graphicx}
\usepackage{indentfirst}
\usepackage{subfigure}
\usepackage{ulem}
\usepackage{bm}
\usepackage{xspace}
\usepackage{threeparttable}
\usepackage{gensymb}
\usepackage{multirow}
\usepackage{epstopdf}
\usepackage{breakurl}
\usepackage{lipsum}
\usepackage[figuresright]{rotating}

\makeatother
\allowdisplaybreaks

\newcommand{\magic}{MAGIC\xspace}
\newcommand{\fermi}{\textit{Fermi}-LAT\xspace}
\newcommand{\argo}{ARGO-YBJ\xspace}
\newcommand{\hess}{H.E.S.S.\xspace}

\newcommand{\ep}{EPWL\xspace}
\newcommand{\lp}{LP\xspace}
\newcommand{\el}{ELP\xspace}
\newcommand{\se}{SEPWL\xspace}

\newcommand{\mrk}{Mrk 421\xspace}
\newcommand{\pks}{PKS 2155$-$304\xspace}
\newcommand{\pg}{PG 1553+113\xspace}

\newcommand{\gev}{\rm GeV\xspace}

\newcommand{\ma}{$m_a$\xspace}
\newcommand{\gag}{$g_{a\gamma}$\xspace}
\newcommand{\gray}{$\gamma$-ray\xspace}

\begin{document}
\title{Searching for Axion-Like Particles with the Blazar Observations\\ of MAGIC and \textit{Fermi}-LAT}
\author{Hai-Jun Li}
\affiliation{Center for Advanced Quantum Studies, Department of Physics, Beijing Normal University, Beijing 100875, China}
\affiliation{Key Laboratory of Particle Astrophysics, Institute of High Energy Physics, Chinese Academy of Sciences, Beijing 100049, China}
\affiliation{School of Physics, University of Chinese Academy of Sciences, Beijing 100049, China}
\author{Xiao-Jun Bi}
\affiliation{Key Laboratory of Particle Astrophysics, Institute of High Energy Physics, Chinese Academy of Sciences, Beijing 100049, China}
\affiliation{School of Physics, University of Chinese Academy of Sciences, Beijing 100049, China}
\author{Peng-Fei Yin}
\affiliation{Key Laboratory of Particle Astrophysics, Institute of High Energy Physics, Chinese Academy of Sciences, Beijing 100049, China}

\date{\today}

\begin{abstract}

We explore the axion-like particle (ALP)-photon oscillation effect in the $\gamma$-ray spectra of the blazars Markarian 421 (Mrk 421) and PG 1553+113, which are measured by the Major Atmospheric Gamma Imaging Cherenkov Telescopes (MAGIC) and \textit{Fermi} Large Area Telescope (\textit{Fermi}-LAT) with high precision.
We investigate the constraints on the ALP parameter space using the Mrk 421 and PG 1553+113 observations of 15 and 5 phases, respectively.
We find that the combined analysis with all the 15 phases improves the limits from the Mrk 421 observations.
The combined limit set by the Mrk 421 observations has excluded the ALP parameter region with the ALP-photon coupling of $g_{a\gamma} \gtrsim 2 \times 10^{-11} \, \rm GeV^{-1}$ for the ALP mass of $\sim 8\times 10^{-9}\, {\rm eV} \lesssim m_a \lesssim 2\times 10^{-7}\rm \, eV$ at 95\% $\rm C.L.$ We also find that the ALP hypothesis can slightly improve the fit to the PG 1553+113 results in some parameter regions, and do not set the limit in this case.


\end{abstract}
\maketitle


\section{Introduction}

The strong \textit{CP} problem is a long standing puzzle in the Standard Model (SM) with the tiny value of $\bar{\theta} \lesssim 10^{-10}$. Introducing an additional spontaneously broken $U(1)$ symmetry, which is also broken by the anomaly at the quantum level, can elegantly solve the strong \textit{CP} problem \cite{Peccei:1977ur, Peccei:1977hh}. This mechanism predicts a light pseudo Nambu-Goldstone boson, called quantum chromodynamics (QCD) axion \cite{Weinberg:1977ma, Wilczek:1977pj}. The axion is also a suitable candidate of cold dark matter. In the early universe, these particles can be nonthermally produced via the misalignment mechanism or the decays of topological defects \cite{Preskill:1982cy, Abbott:1982af, Dine:1982ah, Khlopov:1999tm, Sikivie:2009fv}.

The interactions between the axion and SM particles, such as the photons, leptons, and nucleons, can be described by the effective operators. In the QCD axion models, the axion mass and its couplings to the SM particles are related. From the experimental perspective, searching for the more general parameter space is well motivated. The corresponding particles have the similar effective interactions as the QCD axion, but do not have to solve the strong \textit{CP} problem. Such particles are the so called axion-like particles (ALPs), which are also well motivated in some new physics models beyond the SM, such as the string models \cite{Svrcek:2006yi,Arvanitaki:2009fg,Marsh:2015xka}.

The ALPs have been searched in numerous laboratory and astrophysical experiments for a long time. If the ALP has the coupling to the photons, the ALP and free photon may convert to each other in the external magnetic field \cite{Raffelt:1987im}. The astrophysical magnetic fields on the large scale would induce a detectable ALP-photon oscillation effect.
For the astrophysical source at a large distance, this effect would modify the measured photon spectrum \cite{DeAngelis:2007dqd, Hooper:2007bq}.
In the literature, many studies have been performed to investigate this effect based on the observations of different sources \cite{DeAngelis:2007dqd, Hooper:2007bq, Simet:2007sa,Mirizzi:2007hr,  Mirizzi:2009aj, Belikov:2010ma, Dominguez:2011xy,DeAngelis:2011id,  Horns:2012kw, Abramowski:2013oea,Meyer:2013pny,Mena:2013baa, Tavecchio:2014yoa, Meyer:2014gta,Meyer:2014epa, Reesman:2014ova,  TheFermi-LAT:2016zue, Berenji:2016jji,Meyer:2016wrm,  Kohri:2017ljt,Majumdar:2017vcx,  Galanti:2018upl,Galanti:2018myb,Galanti:2018nvl, Zhang:2018wpc, Liang:2018mqm,Libanov:2019fzq,  Long:2019nrz,   Bi:2020ths, Guo:2020kiq, Buehler:2020qsn, Li:2020pcn, Cheng:2020bhr,Liang:2020roo, Long:2021udi, Davies:2020uxn,  Zhou:2021usu, Batkovic:2021fzr}. Since no ALP effect has been found, these analyses set limits on the ALP mass $m_a$ and the ALP-photon coupling $g_{a\gamma}$ parameter space.

The ALP implication of the high energy \gray spectra of the blazars \pks and \pg, which are measured by \hess and \textit{Fermi} Large Area Telescope (\fermi) \cite{HESS:2016btr} during the common operation time, is investigated in Ref.~\cite{Guo:2020kiq}. The ALP-photon conversion in the turbulent inter-cluster magnetic field $\sim \mathcal{O}(1) \, \mu \rm G$ is considered.
The ALP-photon conversion in the blazar jet magnetic field (BJMF) of Markarian 421 (Mrk 421) is explored in Ref.~\cite{Li:2020pcn}.
The Astrophysical Radiation with Ground-based Observatory at YangBaJing (\argo) and \fermi results covering 10 phases of \mrk \cite{Bartoli:2015cvo} are combined together to set the constraint on the ALP parameter space. Compared with the constraint derived from the individual phase, the combined constraint is significantly improved.

In this work, we use the very high energy (VHE) \gray spectra of Mrk 421 and \pg measured by the Major Atmospheric Gamma Imaging Cherenkov Telescopes (\magic) \cite{Acciari:2019zgl} to investigate the ALP-photon oscillation effect. Compared with Ref.~\cite{Guo:2020kiq}, we study the ALP-photon conversion in the BJMF of \pg in this analysis. Compared with the VHE measurements used in previous studies ~\cite{Guo:2020kiq,Li:2020pcn}, the \magic measurements cover more phases (15 phases for Mrk 421 and 5 phases for \pg) with high precision.
Additionally, the \gray spectra of the blazars at lower energies ($\sim 0.1-100 \rm \, GeV$) can be well constrained by the observation of \fermi. We attempt to combine these results together to search for the ALP-photon oscillation effect and set constraint on the ALP parameter space.

This paper is structured as follows.
In Sec.~\ref{section_alp-p_1}, we briefly introduce the ALP-photon oscillation effect in the VHE astrophysical process
and describe the propagation of the ALP-photon system in the blazar jet, extragalactic space, and Milky Way.
In Sec.~\ref{section_met}, we introduce the data fitting and statistical methods for this analysis.
In Sec.~\ref{section_res}, we investigate the ALP implication in the \magic observations of Mrk 421 and \pg.
The conclusion is given in Sec.~\ref{section_sum}.

\section{The oscillation and prorogation of the ALP-photon system}
\label{section_alp-p_1}

The Lagrangian of ALP including the effective ALP-photon interaction term is
\begin{eqnarray}
\mathcal{L}_{\rm ALP}&=\frac{1}{2}\partial^\mu a\partial_\mu a - \frac{1}{2}m_a^2a^2 -\frac{1}{4}g_{a\gamma}aF_{\mu\nu}\tilde{F}^{\mu\nu},
\end{eqnarray}
where $a$ is the ALP, \ma is its mass, \gag is the coupling between the ALP and photons, and $F_{\mu\nu}$ and $\tilde{F}^{\mu\nu}$ are the electromagnetic field tensor and its dual tensor, respectively.
The ALP-photon system propagating along the $x_3$ direction is written as \cite{DeAngelis:2011id}
$\Psi = \left(A_1, A_2, a \right)^T$, where $A_1$ and $A_2$ denote the linear polarization amplitudes of the photon in the perpendicular directions.
The corresponding density matrix
$\rho=\Psi \otimes \Psi^\dagger$
satisfies the Von Neumann like equation \cite{Mirizzi:2009aj}
\begin{eqnarray}
i\frac{{\rm d}\rho(x_3)}{{\rm d}x_3}=\left[ \rho(x_3), \; \mathcal{M}_0 \right].
\label{lvn}
\end{eqnarray}

Assuming that $B_{\rm T}$ is the transversal magnetic field aligned along the direction of $x_2$, the mixing matrix $\mathcal{M}_0$ can be described by \cite{Raffelt:1987im, Mirizzi:2007hr, Horns:2012kw}
\begin{eqnarray}
\begin{aligned}
\mathcal{M}_0=\begin{pmatrix}
 \Delta_{\rm pl }+2\Delta_{\rm QED }  &  0&0 \\
 0& \Delta_{\rm pl }+\frac{7}{2} \Delta_{\rm QED }  & \Delta_{a \gamma} \\
 0&\Delta_{a\gamma}  & \Delta_{aa}
\end{pmatrix},
\end{aligned}
\end{eqnarray}
with
\begin{eqnarray}
\Delta_{\rm pl}&=&-\frac{\omega_{\rm pl}^2}{2E}\simeq -1.1 \times 10^{-4} \, {\rm kpc}^{-1} \, n_{{\rm cm}^{-3}}E_{\rm GeV}^{-1},\\
\Delta_{\rm QED}&=&\frac{\alpha E}{45\pi} \left( \frac{B_{\rm T}}{B_{\rm cr}}\right)^2 \simeq4.1\times 10^{-9}\, {\rm kpc}^{-1} \, E_{\rm GeV} B_{\mu \rm G}, \ \ \\
\Delta_{a\gamma}&=&\frac{1}{2}g_{a\gamma}B_{\rm T}\simeq 1.52 \times 10^{-2} \, {\rm kpc}^{-1} \, g_{11} B_{\mu \rm G},\\
\Delta_{aa}&=&-\frac{m_a^2}{2E}\simeq -7.8 \times 10^{-2} \, {\rm kpc}^{-1} \, m_{\rm neV}^2 E_{\rm GeV}^{-1},
\end{eqnarray}
where $\omega_{\rm pl}= \sqrt{4\pi \alpha n_e/m_e}$ is the plasma frequency, $n_e$ is the number density of the free electrons, $\alpha$ is the fine-structure constant, and $B_{\rm cr} \equiv m^2_e/|e|\simeq4.4\times 10^{13} \rm \, G$. The terms $\Delta_{\rm pl}$ and $\Delta_{\rm QED}$ represent the plasma and QED vaccum polarisation effects, respectively.
The notations $n_{{\rm cm}^{-3}} \equiv n_e/1 \, \rm {cm}^{-3} $, $E_{\rm GeV} \equiv E /1  \, \rm GeV$, $g_{11}\equiv g_{a\gamma}/10^{-11} \, \rm{GeV}^{-1}$, $B_{\rm \mu G} \equiv B_T /1  \, \mu\rm G$, and $m_{\rm neV} \equiv m_a/1 \, \rm {neV} $ are used in above equations. The general mixing matrix $\mathcal{M}$ depends on the angle $\psi$ between the directions of $B_{\rm T}$ and $x_2$.

The ALP-photon conversion would occur in numerous regions with different magnetic field configurations. The final density matrix can be derived from the solution of Eq.~(\ref{lvn}) as
\begin{eqnarray}
\rho\left(s\right)=T(s)\rho(0)T^\dagger(s).
\end{eqnarray}
The whole transfer matrix $T(s)$ for the propagation distance $s$ reads
\begin{eqnarray}
T(s)=\prod^{n}_{i}\mathcal{T}(i),
\end{eqnarray}
where $\mathcal{T}\left(i\right)$ can be derived from the mixing matrix $\mathcal{M}\left(i\right)$ in the $i$-th region. For the initial unpolarized photon beam with
$\rho(0)=\rm {diag}(1, 1, 0)/2$,
the photon survival probability after propagation is given by \cite{DeAngelis:2011id}
\begin{eqnarray}
P_{\gamma\gamma}={\rm Tr}\left(\left(\rho_{11}+\rho_{22}\right)T(s)\rho(0)T^\dagger(s)\right)
\label{pp}
\end{eqnarray}
with $\rho_{ii}={\rm diag}(\delta_{i1},\delta_{i2},0)$.

Then we describe the prorogation effect of the ALP-photon beam in three astrophysical regions with different magnetic field configurations, including the blazar jet, the extragalactic space, and the Milky Way \cite{Hooper:2007bq,Meyer:2014epa}.
For the BL Lac objects considered in this work, we do not take into account the effects in the blazar broad line region. The ALP-photon oscillation might significantly occur in the BJMF.
There are evidences that the magnetic field of the BL Lac jet can be described by the poloidal (along the jet, reads $B \propto r^{-2}$) and toroidal (perpendicular to the jet, reads $B \propto r^{-1}$) coherent components \cite{Pudritz:2012xj}.
We take the BJMF model of the BL Lac sources as Refs.~\cite{Tavecchio:2014yoa, Galanti:2018upl}.

The transverse magnetic field $B_{\rm jet}(r)$ reads \cite{Begelman:1984mw, Ghisellini:2009wa}
\begin{eqnarray}
B_{\rm jet}(r)=B_0\left(\frac{r}{r_{\rm VHE}}\right)^{-1},
\label{br}
\end{eqnarray}
where $r_{\rm VHE}$ is the distance between the central black hole and emission region.
The density profile of the electrons $n_{\rm el}(r)$ can be given by \cite{OSullivan:2009dsx}
\begin{eqnarray}
n_{\rm el}(r)=n_0\left(\frac{r}{r_{\rm VHE}}\right)^{-2}.
\end{eqnarray}
Note that the above profiles hold in the jet comoving frame. The energies of the photons in the laboratory frame $E_L$ and comoving frame $E_j$ are related by the Doppler factor $\delta_{\rm D}$ through $E_L = E_j\cdot\delta_{\rm D}$.

The fit to the blazar spectra at multi-wave bands with the synchrotron self-Compton model could determine the values of the BJMF parameters. In our analysis, these parameters for one source during all the phases are assumed to be same.
We set $B_0$ to be $0.1\,\rm G$ and $1.0\, \rm G$ for \mrk and \pg, respectively, and take $\delta_{\rm D} = 30$ and $n_0 = 3 \times 10^3 \, \rm {cm}^{-3}$ as the benchmark parameters. These values are consistent with the results derived in Refs.~\cite{Celotti:2007rb,Bartoli:2015cvo}.
In the region with $r > 1\, \rm kpc$, we assume that the magnitude of BJMF is zero. Note that among the BJMF parameters $r_{\rm VHE}$ is difficult to determine through the measurements. Its value might range from $\mathcal{O}(10^{16})$-$\mathcal{O}(10^{17})\,\rm cm$. Here we adopt $r_{\rm VHE}=10^{17}\,\rm cm$ as a benchmark parameter.

When the ALP-photon system propagates in the host galaxy in which the blazar is located, the oscillation effect can be neglected \cite{Tavecchio:2012um, Galanti:2018upl}. If the blazar is located in a cluster with a rich environment, the turbulent inter-cluster magnetic field $\sim \mathcal{O}(1) \, \mu \rm G$ may also induce a significant ALP-photon oscillation effect~\cite{Meyer:2014epa}.
Since no definite evidences that the blazars \mrk and \pg are located in such environment have been provided, this oscillation effect is not considered in our analysis.

The oscillation in the extragalactic magnetic field on the largest cosmological scale is also neglected here. The magnitude of this magnetic field is not larger than $\mathcal{O}(1)\, \rm nG$, while it is not precisely determined already \cite{Ade:2015cva}.
For the VHE photons crossing in the extragalactic space, the attenuation effect caused by the extragalactic background light (EBL) through $\gamma_{\rm VHE} + \gamma_{\rm EBL} \to e^+ + e^-$ should be considered. This effect is described by a suppression factor of $e^{-\tau}$, where $\tau$ is the optical depth depending on the redshift of the source and the EBL density distribution. In this work, we take the EBL model provided by Ref.~\cite{Franceschini:2008tp} as a benchmark.
The redshift of \mrk and \pg are taken as $z_0=0.031$ and 0.45, respectively.

Finally, we take into account the effect in the magnetic field of the Milky Way, where the ALPs could be reconverted to photons. Only the regular component of the Galactic magnetic field is considered here, while the random component on the small scale is neglected. The details of this model can be found in Ref.~\cite{Jansson:2012rt}.

\begin{figure*}[!htbp]
\centering
  \includegraphics[width=1\textwidth]{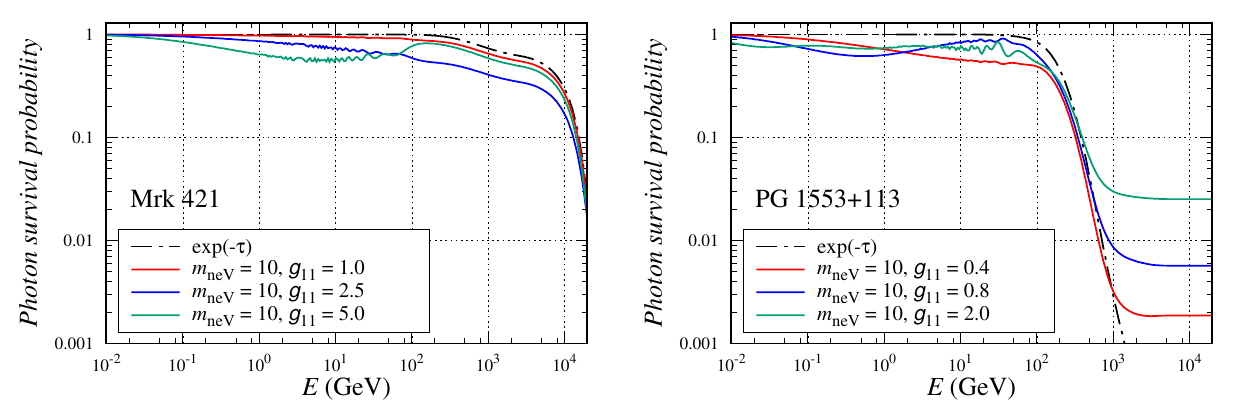}
  \caption{Photon survival probability as a function of energy for \mrk (left) and \pg (right). The black dotted dashed lines represent the survival probability with only the EBL attenuation effect. The solid lines represent the survival probability with both the EBL attenuation and ALP-photon oscillation effects for some selected ALP parameters. The EBL model is taken from Ref.~\cite{Franceschini:2008tp}.}
  \label{fig_prob}
\end{figure*}

We show the photon survival probability $P_{\gamma\gamma}$ as a function of energy for the blazars \mrk and \pg in Fig.~\ref{fig_prob}. It can be seen that the pure EBL attenuation effect described by the factor of $e^{-\tau}$  dramatically suppresses the photon energy spectrum at energies above $\mathcal{O}(10^2)\,\rm GeV$. The ALP-photon oscillation might affect the survival probability at lower energies compared with the EBL attenuation effect. On the other hand, for some ALP parameters, the ALP-photon conversion could compensate the EBL attenuation effect at VHE region and lead to a moderate photon survival probability. This compensation may be significant for \pg at large redshift as shown in Fig.~\ref{fig_prob}.

\section{Gamma-ray data fitting and statistical methods}
\label{section_met}

\magic \cite{Aleksic:2014poa, Aleksic:2014lkm} is a system containing two imaging atmospheric Cherenkov telescopes located at the Roque de los Muchachos Observatory in Spain.
These telescopes could detect extensive air showers in stereoscopic mode, and observe VHE \gray sources at energies above $50\,\rm GeV$ \cite{Acciari:2019zgl}.
In Ref.~\cite{Acciari:2019zgl}, the \magic collaboration reported 32 VHE \gray spectra from 12 blazars. All the data were collected during dark nights in good weather conditions.
The $\gamma$-ray spectra at lower energies $\sim 0.1-100\,\rm GeV$ during the common operation time observed by \fermi are also analyzed in Ref.~\cite{Acciari:2019zgl}. Here we use the \magic results of the BL Lac sources \mrk and \pg covering several activity phases to investigate the ALP-photon oscillation effects.

We take the expressions of the \gray blazar intrinsic energy spectra $\Phi_{\rm int} (E)$ as Ref.~\cite{Acciari:2019zgl}.
$\Phi_{\rm int} (E)$ can be described by some simple functions with three to five parameters, including the power law with exponential cut-off (\ep), power law with superexponential cut-off (\se), log parabola (\lp), and log parabola with exponential cut-off (\el).
The functional expressions of $\Phi_{\rm int} (E)$ are given as follows:
\begin{itemize}
\item  \ep:
\begin{eqnarray}
\Phi_{\rm int} ( E ) = F_0\left(\frac{E}{E_0}\right)^{-\Gamma}\exp\left(-\frac{E}{E_c}\right),
\end{eqnarray}
\item  \se:
\begin{eqnarray}
\Phi_{\rm int} ( E ) = F_0\left(\frac{E}{E_0}\right)^{-\Gamma}\exp\left(-\left(\frac{E}{E_c}\right)^d\right),
\end{eqnarray}
\item \lp:
\begin{eqnarray}
\Phi_{\rm int} ( E ) = F_0\left(\frac{E}{E_0}\right)^{-\Gamma-b \log\left(\frac{E}{E_0}\right)},
\end{eqnarray}
\item \el:
\begin{eqnarray}
\Phi_{\rm int} ( E ) = F_0\left(\frac{E}{E_0}\right)^{-\Gamma-b \log\left(\frac{E}{E_0}\right)}\exp\left(-\frac{E}{E_c}\right),
\end{eqnarray}
\end{itemize}
where $F_0$, $E_c$, $\Gamma$, $b$, and $d$ are free parameters. For \ep and \se, $E_0$ is taken to be 1 \gev, while for \lp and \el, $E_0$ is also treated as a free parameter.
For each phase, we choose the intrinsic energy spectrum with the minimum best-fit reduced $\chi^2$ under the null hypothesis. This is different from the analysis in Ref.~\cite{Li:2020pcn} where
the expression of the intrinsic energy spectrum is same for all the phases.
The spectrum expressions for all the phases adopted in this analysis are listed in Table~\ref{tab_1}.

\begin{table*}
\caption{The best-fit values of ${\chi}_{\rm w/oALP}^2$ under the null hypothesis and $\chi^2_{\rm min}$ under the ALP hypothesis for all the phases. Periods stand for the corresponding MAGIC observations. The expressions of the intrinsic energy spectra, the effective $\rm d.o.f.$ of the TS distributions, and $\Delta{\chi}^2$ at 95\% $\rm C.L.$ are also listed. The last two rows denote the results of the combined analysis.}
\begin{ruledtabular}
\begin{tabular}{lccccccr}
Source [period]    &     Tstart    &   Tstop    &   Spectrum  &  ${\chi}_{\rm w/oALP}^2$ &     $\chi^2_{\rm min}$    &   Effective d.o.f.    &   $\Delta\chi^2$    \\
\hline
\mrk [20130410]   &    2013-04-09T12:00 & 2013-04-10T12:00   &  \se     &  12.244        &   8.845    &  4.45   &   10.232  \\
\mrk [20130411]   &    2013-04-10T18:00 & 2013-04-11T06:00   &  \el     &  16.213        &   10.124   &  6.85   &   13.868  \\
\mrk [20130412]   &    2013-04-11T18:00 & 2013-04-12T06:00   &  \el     &  8.911         &   6.186    &  7.54   &   14.868  \\
\mrk [20130413a]  &    2013-04-12T12:00 & 2013-04-13T12:00   &  \el     &  16.007        &   12.928   &  8.00   &   15.527 \\
\mrk [20130413b]  &    2013-04-12T12:00 & 2013-04-13T12:00   &  \se     &  9.733         &   8.645    &  4.72   &   10.657    \\
\mrk [20130413c]  &    2013-04-12T12:00 & 2013-04-13T12:00   &  \se     &  10.049        &   7.537    &  4.56   &   10.406    \\
\mrk [20130414]   &    2013-04-13T12:00 & 2013-04-14T12:00   &  \el     &  22.391        &   13.749   &  9.22   &   17.245   \\
\mrk [20130415a]  &    2013-04-14T21:17 & 2013-04-15T04:13   &  \el     &  5.774         &   4.777    &  5.23   &   11.447    \\
\mrk [20130415b]  &    2013-04-14T21:17 & 2013-04-15T04:13   &  \se     &  13.426        &   10.016   &  5.02   &   11.124 \\
\mrk [20130415c]  &    2013-04-14T21:17 & 2013-04-15T04:13   &  \se     &  5.056         &   4.012    &  4.71   &   10.641   \\
\mrk [20130416]   &    2013-04-15T12:00 & 2013-04-16T09:00   &  \se     &  32.863        &   19.552   &  5.48   &   11.829     \\
\mrk [20130417]   &    2013-04-16T18:00 & 2013-04-17T06:00   &  \se     &  26.050        &   11.174   &  4.99   &   11.077  \\
\mrk [20130418]   &    2013-04-17T12:00 & 2013-04-18T12:00   &  \ep     &  13.345        &   9.038    &  5.56   &   11.950  \\
\mrk [20130419]   &    2013-04-18T12:00 & 2013-04-19T12:00   &  \el     &  3.609         &   1.964    &  4.07   &    9.625   \\
\mrk [20140426]   &    2014-04-25T18:00 & 2014-04-26T06:00   &  \el     &  25.809        &   15.184   &  6.19   &   12.896  \\
\hline
\pg [ST0202]      &    2012-02-28T12:00 & 2012-03-04T12:00   &  \ep     &  2.326         &   0.914    &  3.52   &    8.723   \\
\pg [ST0203]      &    2012-03-13T12:00 & 2012-05-02T12:00   &  \se     &  15.598        &   6.342    &  6.24   &   12.970 \\
\pg [ST0302]      &    2013-04-07T12:00 & 2013-06-12T12:00   &  \se     &  5.413         &   1.279    &  4.50   &    10.311   \\
\pg [ST0303]      &    2014-03-11T12:00 & 2014-03-25T12:00   &  \ep     &  10.171        &   5.944    &  6.73   &   13.693    \\
\pg [ST0306]      &    2015-01-25T12:00 & 2015-08-07T12:00   &  \se     &  4.704         &   0.718    &  3.43   &    8.578   \\
\hline
Combined \mrk     &                    &                     &          &  221.480       &   204.554  &  31.17  &   45.206   \\
Combined \pg      &                    &                     &          &  38.212        &   20.511   &  8.94   &   16.854    \\
\end{tabular}
\end{ruledtabular}
\label{tab_1}
\end{table*}

Under the alternative hypothesis including the ALP-photon oscillation effect, we obtain the expected photon spectrum as
\begin{eqnarray}
\Phi_{\rm w \; ALP} ( E ) = P_{\gamma\gamma} \Phi_{\rm int} ( E ),
\end{eqnarray}
where $P_{\gamma\gamma}$ is the photon survival probability.
The detected photon flux in the energy bin of $(E_1,E_2)$ is given by \cite{Guo:2020kiq, Li:2020pcn}
\begin{eqnarray}
\Phi'=\frac{\int_0^{\infty}D(E', E_1, E_2)\Phi ( E' ){\rm d}E'}{E_2 - E_1},
\end{eqnarray}
where $D(E', E_1, E_2)$ is the energy dispersion function, and $E'$ and $\Phi(E')$ are the energy and spectrum of the photons before detection, respectively. The energy resolution of \magic is taken to be 16\% \cite{Aleksic:2014lkm}.

In Ref.~\cite{Acciari:2019zgl}, the \fermi spectra are provided in the form of spectral bow-ties rather spectral points.
The bow-ties contain the information of the flux and local spectrum index determined at the decorrelation energy; each one contributes two degrees of freedom in the fit.
The $\chi^2$ of the fit is defined as \cite{Acciari:2019zgl}
\begin{eqnarray}
\begin{aligned}
\chi^2 &= \left(\frac{\Phi'(E_{\rm LAT})-F_{\rm LAT}}{\Delta F_{\rm LAT}}\right)^2+ \left(\frac{\Gamma_{\rm fit}-\Gamma_{\rm LAT}}{\Delta \Gamma_{\rm LAT}}\right)^2 \\
&+ \sum_{i=1}^{N} \left(\frac{\Phi'(E_i) - \tilde{\phi}_i}{\delta_i}\right)^2,
\end{aligned}
\end{eqnarray}
where $E_{\rm LAT}$, $F_{\rm LAT}$, $\Gamma_{\rm LAT}$, and $\Gamma_{\rm fit}$ are the central energy, flux, local spectral index, and the expected spectral index for the \fermi results, respectively.
$N$ is the number of the \magic spectral points, $\Phi'(E_i)$ is the expected flux of the photons, $\tilde{\phi}_i$ is the detected photon flux, and $\delta_i$ is the uncertainty of the \magic measurement.

With the $\chi^2$ values under the ALP hypothesis in the $m_a-g_{a\gamma}$ plane, the constraint on the parameter space is set by requiring $\chi^2 \leq \chi_{\rm min}^2 + \Delta{\chi}^2$,
where ${\chi}_{\rm min}^2$ is the minimum best-fit ${\chi}^2$ under the ALP hypothesis.
Since the modifications of the balzar spectra nonlinearly depend on the ALP parameters, the threshold value of $\Delta{\chi}^2$ at the particular confidence level should be derived from the Monte Carlo simulations rather than directly using Wilks' theorem \cite{Meyer:2014epa,TheFermi-LAT:2016zue}.
Based on the best-fit spectra to the data under the null hypothesis, for each phase, 400 sets of spectra in the pseudo-experiments are generated by Gaussian samplings.
The test statistic (TS) value is defined by the difference between the best-fit ${\widehat{\chi}}^2$ under the null and ALP hypotheses for each generated spectrum ${\rm TS}\equiv {\widehat{\chi}_{\rm null}}^2 - {\widehat{\chi}_{\rm w \; ALP}}^2$.
In each phase, the distribution of TS for all the generated spectrum sets is derived. Such distribution can be described by the non-central $\chi^2$ distribution with the non-centrality $\lambda$ and the effective degree of freedom ($\rm d.o.f.$). Although this TS distribution is derived under the null hypothesis, following Ref.~\cite{TheFermi-LAT:2016zue} we take it as the approximation of the TS distribution under the ALP hypothesis and adopt the corresponding $\Delta{\chi}^2$ in the following analysis.

\section{Constraints on the ALP parameter space}
\label{section_res}

\begin{figure*}[!htbp]
\includegraphics[width=1\textwidth]{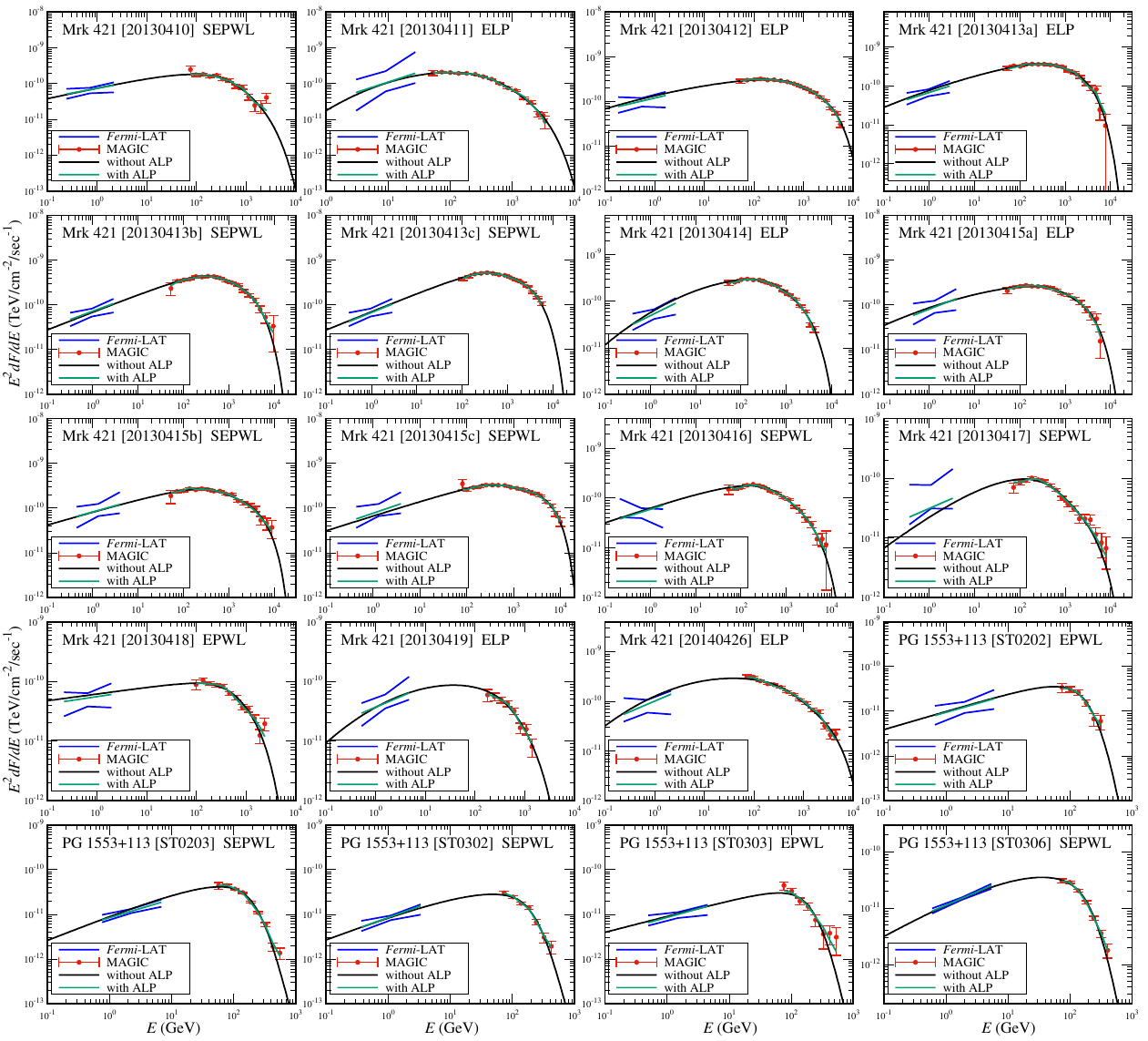}
\caption{Best-fit photon spectra for the 15 and 5 phases of \mrk and \pg , respectively.
The black and green lines represent the spectra under the null and ALP hypotheses, respectively.
The values of the corresponding best-fit $\chi^2$ are listed in Table~\ref{tab_1}.
The spectral points and bow-ties represent the results from \magic and \fermi \cite{Acciari:2019zgl}, respectively.}
\label{fig_dfde}
\end{figure*}

\begin{figure*}[!htbp]
\includegraphics[width=1\textwidth]{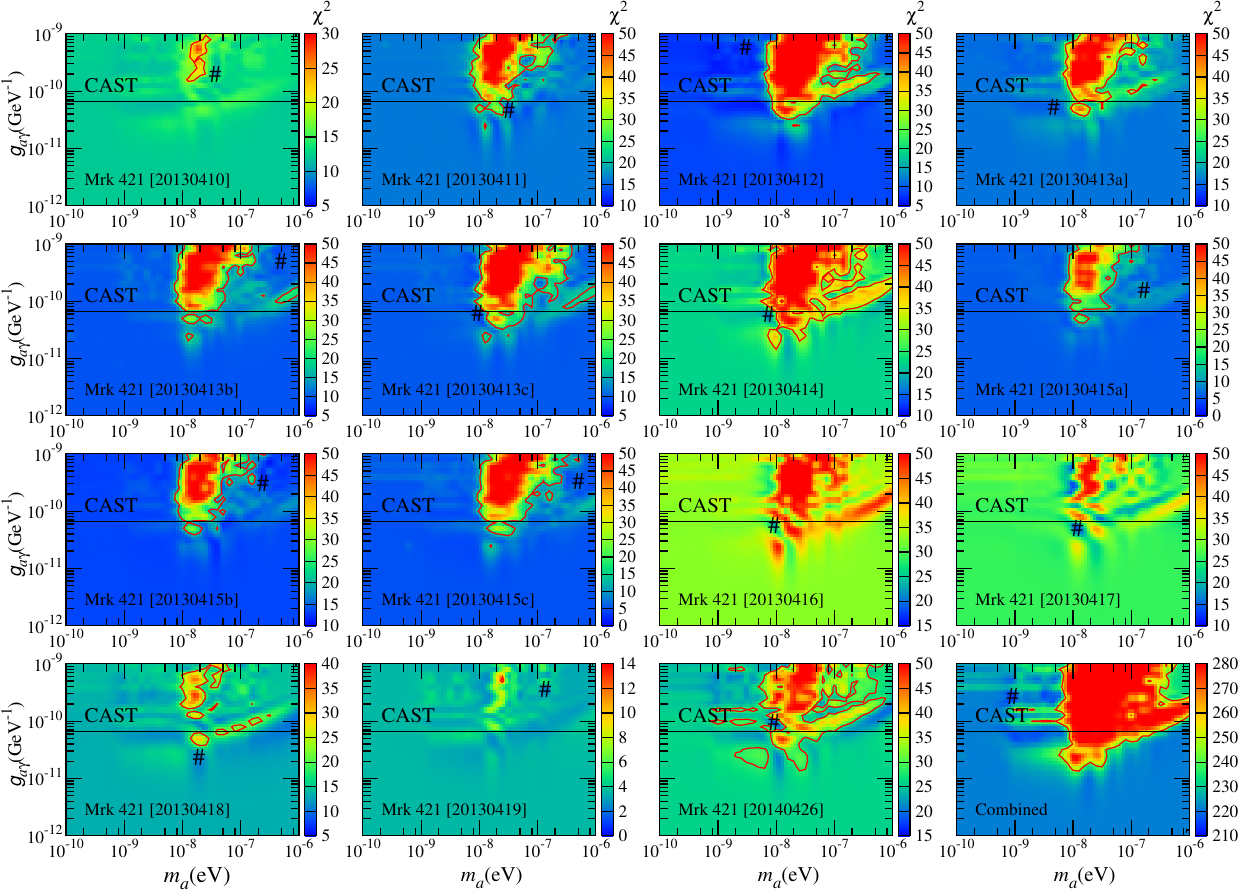}
\caption{$\chi_{\rm w \; ALP}^2$ values in the $m_a-g_{a\gamma}$ plane for the 15 phases of \mrk. The $\chi_{\rm w \; ALP}^2$ values for the combined results are shown in the bottom right panel.
The red contours represent the excluded regions at 95\% $\rm C.L.$ The ``\#" symbols represent the best-fit ALP parameter points.
The horizontal line represents the upper limit placed by CAST \cite{Anastassopoulos:2017ftl}.}
\label{fig_chi2_1}
\end{figure*}

In this section, we investigate the implication of ALP for the observations of \magic and \fermi. The best-fit ${\chi}_{\rm w/oALP}^2$ under the null hypothesis and $\chi_{\rm w \; ALP}^2$ under the ALP hypothesis are given by Table~\ref{tab_1}.
We calculate the TS distributions for all the phases and obtain
their non-centralities $\sim 0.01$.
The corresponding effective $\rm d.o.f.$ and the values of $\Delta{\chi}^2$ at $95\%$ $\rm C.L.$ are also given by Table~\ref{tab_1}.

The best-fit photon spectra under the null and ALP hypotheses for all the phases are shown in Fig.~\ref{fig_dfde}.
We find that the null hypothesis can well fit the \mrk observations. The corresponding best-fit reduced $\chi^2$ are around an average value of 1.10. For the most of phases, introducing the ALP-photon oscillation would not significantly improve the fit.
With the values of $\Delta{\chi}^2$, the constraints on the ALP parameter space at 95\% $\rm C.L.$ from the \mrk observations are represented by the red contours in Fig.~\ref{fig_chi2_1}.
We find that not all the observations of the single phase can be used to set the 95\% $\rm C.L.$ constraint on the ALP parameter space.
Following Ref.~\cite{Li:2020pcn}, we also perform an analysis combined the \mrk results of the 15 phases. This approach could give a more reliable implication.
The combined $\chi_{\rm w \; ALP}^2$  in the $m_a-g_{a\gamma}$ plane and the best-fit value are shown in Fig.~\ref{fig_chi2_1} and Table~\ref{tab_1}, respectively.
The red contour representing the combined upper limit at 95\% $\rm C.L.$ is also shown.

\begin{figure*}[htb]
\centering
\includegraphics[width=0.75\textwidth]{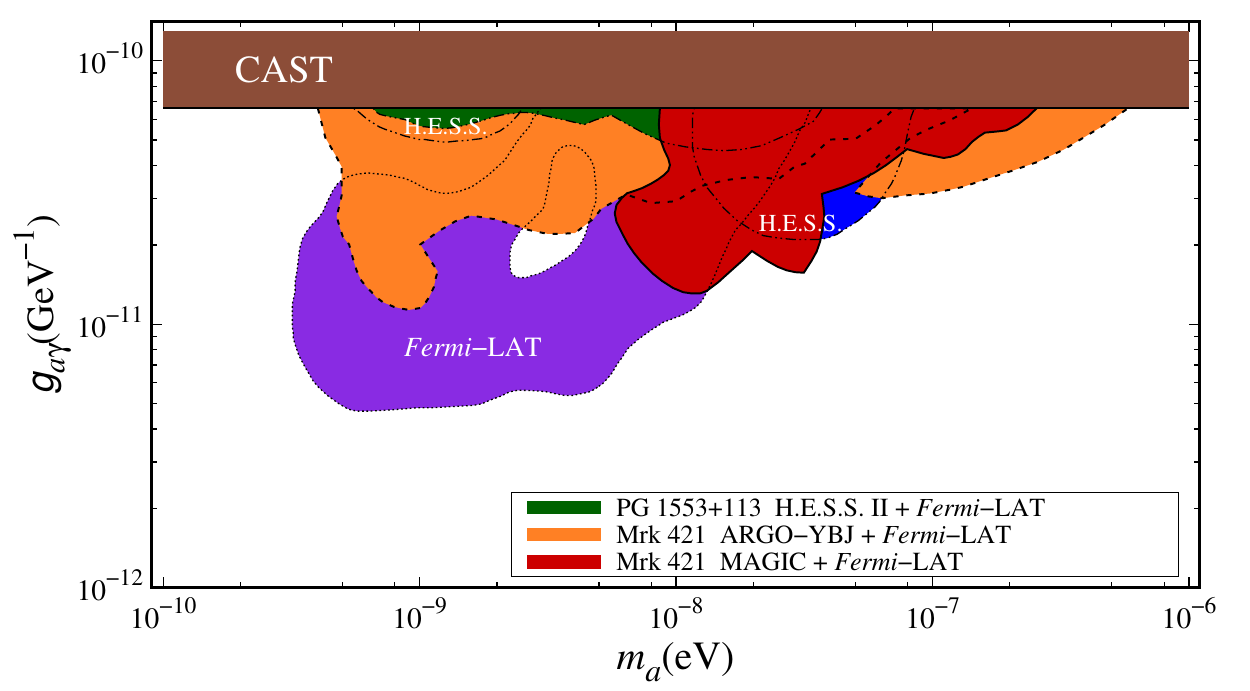}
\caption{95\% $\rm C.L.$ upper limit (red contour) placed by the \mrk observations of \magic and \fermi.
The upper limits set by CAST \cite{Anastassopoulos:2017ftl}, the \pks observation of \hess \cite{Abramowski:2013oea}, and the NGC 1275 observation of \fermi \cite{TheFermi-LAT:2016zue} are shown for comparison.
The limits placed by the analyses using the \mrk observations of \argo and \fermi ~\cite{Li:2020pcn}, and the \pg observations of $\rm \hess~II$ and \fermi~\cite{Guo:2020kiq} are also shown.
}
\label{fig_chi2_com}
\end{figure*}

In Fig.~\ref{fig_chi2_com}, the constraints on the ALP parameter space placed by CAST \cite{Anastassopoulos:2017ftl}, the \pks observation of \hess \cite{Abramowski:2013oea}, and the NGC 1275 observation of \fermi \cite{TheFermi-LAT:2016zue} are shown for comparison. We also show the limits set by the analyses using the \mrk observations of \argo and \fermi ~\cite{Li:2020pcn}, and the \pg observations of $\rm \hess~II$ and \fermi~\cite{Guo:2020kiq} in Fig.~\ref{fig_chi2_com}.
Compared with the CAST constraint of $ g_{a\gamma} \lesssim 6.6\times 10^{-11}\, \rm GeV^{-1}$ \cite{Anastassopoulos:2017ftl}, the combined limit  at 95\% $\rm C.L.$ set by this work excludes the ALP parameter region with the ALP-photon coupling of $g_{a\gamma} \gtrsim 2 \times 10^{-11}\, \rm GeV^{-1}$ for the ALP mass of $\sim 8\times 10^{-9}\, {\rm eV} \lesssim m_a \lesssim 2\times 10^{-7}\, \rm eV$.
This combined constraint is not completely coincide with that derived from the observations of \argo and \fermi in Ref.~\cite{Li:2020pcn}.
A possible reason is that the spectral forms of the \fermi results are different in these two analyses.
The spectral points of the \fermi result provide a large contributions to the final $\chi^2$ in Ref.~\cite{Li:2020pcn}. On the other hand, the \fermi results used in this analysis are in the form of bow-ties with two parameters. Therefore, the VHE data from \magic would provide the dominant contributions to the final $\chi^2$ in this analysis.
Additionally, the intrinsic energy spectra for all the phases are assumed to be same in Ref.~\cite{Li:2020pcn}, while they are separately chosen according the fits for different phases in this analysis. This difference would also induce different fitting results.

\begin{figure*}[!htbp]
\includegraphics[width=0.75\textwidth]{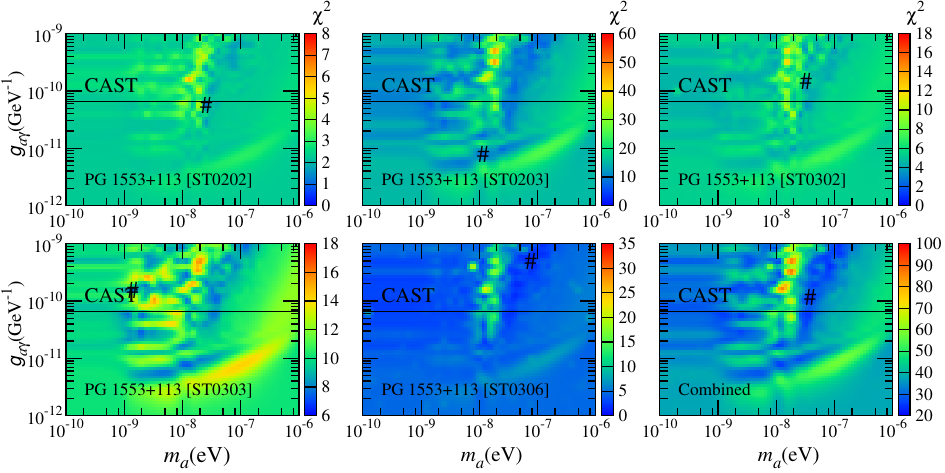}
\caption{$\chi_{\rm w \; ALP}^2$ values in the $m_a-g_{a\gamma}$ plane for the 5 phases of \pg. The $\chi_{\rm w \; ALP}^2$ values for the combined results are shown in the bottom right panel. The ``\#" symbols represent the best-fit ALP parameter points.
The horizontal line represents the upper limit placed by CAST \cite{Anastassopoulos:2017ftl}.}
\label{fig_chi2_2}
\end{figure*}

For \pg, the best-fit reduced $\chi^2$ are around an average value of 1.23. For the phases except \pg [ST0203], the ALP hypothesis does not significantly improve the fit. However, for \pg [ST0203], the difference between the best-fit $\chi^2$ values under the null and ALP hypotheses is near the threshold $\Delta \chi^2$ at 95\% $\rm C.L.$ as shown in Table~\ref{tab_1}. Combining all results of the 5 phases, we find that this difference becomes larger than $\Delta \chi^2$ at 95\% $\rm C.L.$ In this case, we only show the values of $\chi^2$ in the $m_a-g_{a\gamma}$ plane in Fig.~\ref{fig_chi2_2}, but do not set the constraints on the ALP parameter space.

Some comments on these results are given as follows. The ALP-photon oscillation effect strongly depends on the magnitude of the astrophysical magnetic field. For \pg, we take a relative large value of $B_0=1\, \rm G$, which directly enhances the oscillation effect. Since the ALPs do not interact with the EBL, the large oscillation effect could compensate the attenuation effect and reduce the absorption of the VHE photons in the extragalactic space, especially for the astrophysical source at large redshift suffering from a significant attenuation effect. Therefore, for \pg at $z_0\sim 0.45$, the oscillation effect might induce a relative large photon flux at VHE band compared with the null hypothesis. From Fig.~\ref{fig_dfde}, we can see that the ALP hypothesis improves the fit to the last one or two data points of the \magic measurements, which seem not to drop dramatically compared with the perivenous data points. This behavior can be explicitly seen in the spectrum of the phase \pg [ST0303], despite the uncertainties of this phase are relative large. On the other hand, the spectrum of the phase \pg [ST0203] has small uncertainties and could be used to reveal the oscillation effect.

We emphasize that the results discussed above are affected by the astrophysical uncertainties. The dominant uncertainties are from the BJMF model.
In the BJMF model used in this work, the magnitude of the magnetic field depends on the parameters $B_0$, $\delta_{\rm D}$, $n_0$, and $r_{\rm VHE}$. As discussed in Refs.~\cite{Meyer:2014epa, Li:2020pcn}, the distance between the VHE emission site and the central black hole $r_{\rm VHE}$, and the magnitude of the core magnetic field $B_0$ at $r_{\rm VHE}$ would significantly affect final results.
The parameter $B_0$ directly characterizes the magnitude of the BJMF. In principle, this parameter can be obtained from the fit to the blazar spectrum using the synchrotron self-Compton model. However, the value of $r_{\rm VHE}$ is difficult to precisely determine.

\begin{figure*}[!htbp]
\includegraphics[width=0.75\textwidth]{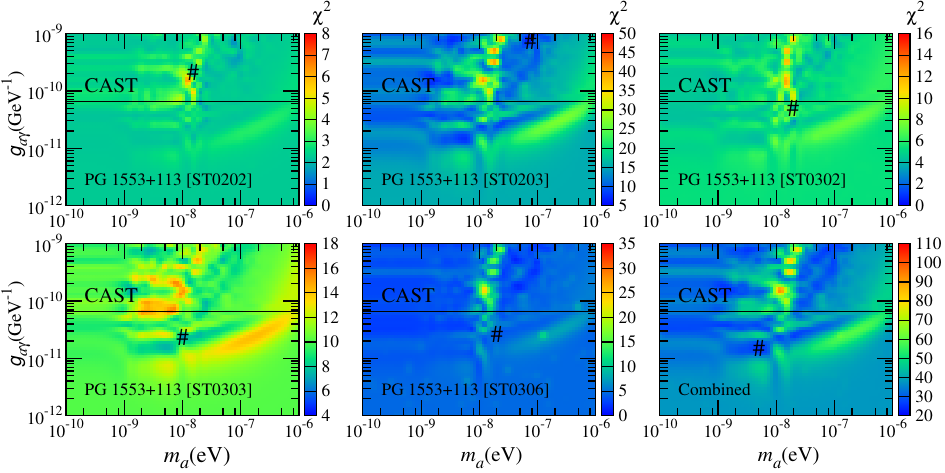}
\caption{Same as Fig.~\ref{fig_chi2_2} but for $r_{\rm VHE}=3 \times 10^{16}\, \rm cm$.}
\label{fig_chi2_3}
\end{figure*}

As discussed in Ref.~\cite{Li:2020pcn}, for the increasing value of $r_{\rm VHE}$ in the range of $\sim 10^{16}-10^{18}\, \rm cm$, the final constraint from the \mrk observations would also become more strict by a magnitude of $1-2$ orders. The results for \mrk in this analysis have similar dependence on $r_{\rm VHE}$. For \pg, we perform an analysis for a small value of $r_{\rm VHE}$ as $3\times 10^{16}\, \rm cm$. We find that the difference between the best-fit $\chi^2$ under the null and ALP hypotheses is 17.068, which is slightly larger than the threshold value at 95\% C.L. of 16.587. In this case, since the ALP hypotheses is able to improve the fit, the constraint on the parameter space is not set. The corresponding $\chi^2$ under the ALP hypothesis for \pg are shown in Fig.~\ref{fig_chi2_3}. We can see that for the fixed $m_a$, the behavior of the change of $\chi^2$ for $r_{\rm VHE}=3\times 10^{16}\, \rm cm$ is similar with that at smaller $g_{a\gamma}$ for $r_{\rm VHE}=10^{17}\, \rm cm$.

\section{Conclusion}
\label{section_sum}

In this work, we analyze the ALP-photon oscillation effect in the spectra of the blazars \mrk and \pg measured by \magic and \fermi during the common operation time, that covers the 15 and 5 activity phases, respectively.
We find that not all the observations of these phases can be individually used to set the 95\% $\rm C.L.$ limit on the ALP parameter space.
For \mrk, we find that the constraint can be significantly improved if the results of all the 15 phases are combined.
The combined \mrk observations of \magic and \fermi have excluded the ALP parameter region with the ALP-photon coupling of $g_{a\gamma} \gtrsim 2 \times 10^{-11}\, \rm GeV^{-1}$ for the ALP mass of $\sim 8\times 10^{-9}\, {\rm eV} \lesssim m_a \lesssim 2\times 10^{-7}\, \rm eV$ at 95\% $\rm C.L.$
For \pg, we find that the ALP hypothesis can slightly improve the fit to the data in some parameter regions. However, since the anomalies of the intrinsic spectrum and the EBL model may also induce the similar effect, we do not make a further ALP interpretation for the current observation.

In the future, the new generation VHE $\gamma$-ray observations, such as Cherenkov Telescope Array \cite{Acharya:2013sxa}, Large High Altitude Air Shower Observatory \cite{Cao:2010zz}, High Energy cosmic-Radiation Detection \cite{Huang:2015fca}, Gamma-Astronomy Multifunction Modules Apparatus \cite{Egorov:2020cmx}, and Tunka Advanced Instrument for Gamma-ray and Cosmic ray Astrophysics-Hundred Square km Cosmic ORigin Explorer \cite{Kuzmichev:2018mjq}, will collect more data for the high energy $\gamma$-ray sources at large distances from the Earth with high precision.
With these precise $\gamma$-ray observations for several blazars, it is possible to test the ALP-photon oscillation at the VHE band or set the more stringent constraints on the ALP parameters.

\section*{Acknowledgments}
The authors would like to thank Mireia Nievas Rosillo for providing the energy spectra of \mrk and \pg measured by \magic and \fermi in the common operation time.
We also thank Jun-Guang Guo for providing helpful discussions and comments.
This work is supported by the National Key R\&D Program of China (Grant No.~2016YFA0400200) and the National Natural Science Foundation of China (Grants No.~U1738209 and No.~11851303).

\bibliography{ref}

\end{document}